\documentstyle [12pt,epsf]{article}
\def\gev{{\rm GeV}}
\def\mev{{\rm MeV}}

\newcommand{\beq}{\begin{equation}}
\newcommand{\eeq}{\end{equation}}
\newcommand{\bea}{\begin{eqnarray}}
\newcommand{\eea}{\end{eqnarray}}
\newcommand{\bsub}{\begin{subequations}}
\newcommand{\esub}{\end{subequations} \noindent}

\catcode`\@=11 
\def\lsim{\mathrel{\mathpalette\@versim<}}
\def\gsim{\mathrel{\mathpalette\@versim>}}
\def\@versim#1#2{\vcenter{\offinterlineskip
        \ialign{$\m@th#1\hfil##\hfil$\crcr#2\crcr\sim\crcr } }}
\catcode`\@=12 
\begin{document}
\begin{titlepage}
\begin{center}
    \vspace*{1.2cm}
       {\Large\bf Can Symmetric Texture reproduce \\
         Neutrino Bi-large Mixing?}
    \vspace{1.5cm} \\
        {\large
      Masako {\sc Bando $^1$} \footnote{E-mail address:
        bando@aichi-u.ac.jp},  
      Satoru {\sc Kaneko  $^2$} \footnote{E-mail address:
        kaneko@muse.sc.niigata-u.ac.jp},
      Midori {\sc Obara $^3$} \footnote{E-mail address:
        midori@hep.phys.ocha.ac.jp} \\
   and \\
  Morimitu {\sc Tanimoto  $^2$} \footnote{E-mail address:
        tanimoto@muse.sc..niigata-u.ac.jp}}   \\
    \vspace{7mm}
    $^1$ {\it Aichi University, Aichi 470-0296, Japan} \\[1mm]
    $^2$  {\it Department of Physics,
       Niigata University, Niigata,  950-2128, Japan}  \\[1mm] 
    $^3$ {\it Institute  of Humanities and Sciences, \\ 
    Ochanomizu University, Tokyo 112-8610, Japan}  
\end{center}
    \vspace{0.5cm}
\begin{abstract}
We study the symmetric texture of geometric form with 2-zeros to see if 
it is consistent with the presently-known neutrino masses and mixings. 
In the neutrino mass matrix elements
we obtain numerically the allowed region of the parameters 
including CP violating phases, which can reproduce the present neutrino 
experiment data. 
The result of this analysis dictates
the narrow  region for the GUT model including Pati-Salam symmetry 
with texture zeros to be consistent 
with the experimental data. The $|U_{e3}|$ and $J_{CP}$ are 
 also predicted in such models.
\end{abstract}
\end{titlepage}
\setcounter{footnote}{0}

Neutrino experiments by Super-Kamiokande  
\cite{SKam,SKamsolar} 
 and SNO\cite{SNO} 
have brought us an outstanding fact on  
the neutrino oscillation.
  Recent results from  KamLAND  have almost confirmed  the large neutrino
 mixing solution that is responsible for the solar neutrino problem
nearly uniquely \cite{KamLAND}.  
We have now common information concerning the neutrino
mass difference squared ($\Delta m^2_{\rm atm}$, $\Delta m^2_{\rm sun}$) and 
neutrino flavor mixings ($\sin^2 2\theta_{\rm atm}$ and 
 $\tan^2 \theta_{\rm sun}$) \cite{Lisi}
as follows:
\begin{eqnarray}
 0.35 \le  \tan^2 \theta_{12} \le 0.54  , \ \ 
&&6.1\times 10^{-5} \le  \Delta m^2_{\rm sun} \le  
8.3\times 10^{-5}~\rm{eV}^2,  
 \quad 90\% C.L. \nonumber \\ 
0.90 \le \sin^2 2\theta_{23}  , \ \ 
&& 1.3\times 10^{-3} \le  \Delta m^2_{\rm atm} \le 3.0 \times 10^{-3}~
 \rm{eV}^2 ,  \quad 90\% C.L. \ .
\label{expangle}
\end{eqnarray}
In these data it is remarked  that the  neutrino mixing  is the bi-large
 and the ratio $ \Delta m^2_{\rm sun}/ \Delta m^2_{\rm atm} $
is $\sim \lambda^2$ with $\lambda\simeq 0.2$.
 A  constraint has also been placed on the third mixing angle  
from the reactor experiment of CHOOZ \cite{CHOOZ}.
 These results  are  very important for model buildings of flavors.

There are many attracting points in grand unified theories (GUT), 
anomaly cancellation between quarks and leptons in one family, 
gauge coupling unification, electromagnetic charge quantization, e.t.c.. 
In the framework of GUT, quarks and leptons are unified in some way and 
their masses and mixing angles are mutually related. 
Now  
the neutrino sector which shows less hierarchical and bi-large 
mixing angles is quite different from the quark sector 
where far stronger hierarchy is observed 
with very tiny mixing angles. 
So the problem is whether such large difference of 
quark and lepton sectors can 
be consistent with GUT. 
So far as we assume general $U(1)$ family 
structure \cite{bandokugopr} with order $1$ 
coefficients of Yukawa couplings, 
the simplest example of symmetric mass matrix 
is already excluded because the resultant neutrino mass matrix 
is predicted to be also  hierarchical with small mixings. 
 However if we assume some additional symmetry to protect some components 
of the mass matrix  leading  "zero" texture, 
the above statement is no more guaranteed \cite{Bando}. 
Actually in the previous paper \cite{BaOba} an example of 
symmetric 4 zero texture is shown to reproduce the bi-large 
neutrino mixing  compatible with GUT. 
On the other hand, the experimental 
data already dictates the desired form of neutrino mass matrix $M_{\nu}$ 
for which the order of each component is as follows \cite{uibandokugo};  
\begin{eqnarray}
M_{\nu}\sim
\left(
\begin{array}{@{\,}ccc@{\,}}
\lambda^2     & \lambda   & \lambda    \\
\lambda   & 1    & 1 \\
 \lambda & 1 & 1
\end{array}
\right) m_{\nu} \ .  
\label{generalordermass}
\end{eqnarray}
Note that,  in order for the above form to reproduce the bi-large mixing 
with  the observed mass squared differences, 
it is not sufficient to discuss only the order of magnitudes, and
we have to tune the coefficients very carefully. 
The minimum texture preserving the above properties 
would be the one having some zeros \cite{Fu,Fram,Xing,KKST,Barbieri}, where 
we need the 23 element of order 1 to get large 23 mixing 
angle, and further the determinant of the $2\times 2$ matrix of 
the right bottom corner should become of order $\lambda$ 
in order to reproduce the experimental mass difference ratio 
$\Delta m^2_{\rm sun}/\Delta m^2_{\rm atm}$, 
the 22 element should be of order 1. 
Also the 12 (13)  element  must be non-zero to reproduce large mixing 
angle $\theta_{12}$. 
So the only possible zeros are for  11 and 13 (12) elements 
namely two-zero symmetric texture. 
Thus we can take the simplest form of neutrino mass matrix at GUT scale 
as a  minimal model 
\footnote{Another 2-zero texture has been adopted by Chen and Mahanthappa
\cite{CM}.}
including a phase $\phi$; 
\begin{eqnarray}
M_{\nu}=
 m_{\nu} \left(
\begin{array}{@{\,}ccc@{\,}}
 0               & \beta  & 0   \\
\bar \beta  & \bar \alpha   & \bar h \\
 0  & \bar h & 1
\end{array}
\right)
=
 m_{\nu} \ P_{\nu}^{T}
\left(
\begin{array}{@{\,}ccc@{\,}}
 0               & \beta  & 0   \\
\beta  & e^{i\phi}\alpha   & h \\
 0  & h & 1
\end{array}
\right)P_{\nu}\ , 
\quad 
\begin{array}{@{\,}c@{\,}}
  \beta \simeq O(\lambda)         \\
 \alpha  \simeq O(1)        \\
  h \simeq O(1)
\end{array}\ , 
\label{generalform}
\end{eqnarray}
with $\bar\alpha, \bar\beta,\bar h$, 
being made  positive real numbers, 
  $\alpha, \beta, h$ by factored out the phases by the diagonal phase
matrix $P_{\nu}$
\footnote{This kind of 4-zero case has been 
studied extensively for the quark masses;  
\begin{eqnarray}
M_u=
\left(
\begin{array}{@{\,}ccc@{\,}}
 0   & A  & 0   \\
A  & B   & C \\
 0  & C & 1
\end{array}
\right)m_t \ , \quad 
M_d=
\left(
\begin{array}{@{\,}ccc@{\,}}
 0    & A'  & 0   \\
A'  & B'   & C' \\
 0  & C' & 1
\end{array}
\right)m_b \ .  \nonumber
\end{eqnarray}
Here the matrix is assumed to be factored out 
by $P$ in the four-zero texture case, which is exactly 
possible in the case of 6-zero texture. 
Note that we cannot factor out all the phases 
to make the matrix elements of $M$ all real and there remains 
one phase  as is seen in Eq.~(\ref{generalform}).}.

In this letter we investigate this kind of 2-zero  texture 
including CP phase and examine  
 parameter regions which are consistent with 
the present experiments. 
The neutrino and quark mixings 
are expressed by MNS~\cite{MNS}  and  CKM  matrices, respectively, 
\begin{equation}
U_{MNS}=U_l^{\dagger} U_{\nu}\ ,\qquad U_{CKM}=U_u^{\dagger} U_{d}\ , 
\label{CKMNS}
\end{equation}
which are 
further divided into two unitary matrices, $U_u$ and $U_d$ 
or  $U_l$ and $U_{\nu}$, respectively, 
which diagonalize 
the $3\times 3$ up and down quark mass matrices  $M_u$ and $M_d$
or charged lepton and neutrino mass matrices, $M_l$ and $M_{\nu}$ 
respectively; 
\begin{eqnarray}
U_l^{\dagger}M_lV_l&=&{\rm diag}(m_e,m_{\mu},m_{\tau}), \quad 
U_{\nu}^{T}M_{\nu}U_{\nu}={\rm diag}(m_{\nu_1},m_{\nu_2},m_{\nu_3}), 
\\
U_u^{\dagger}M_uV_u &=& 
{\rm diag}(m_u,m_c,m_t),  \quad 
U_d^{\dagger}M_dV_d = 
{\rm diag}(m_d,m_s,m_b), 
\label{unitary}
\end{eqnarray}
where $U$ and $V$ are unitary matrix acting on left- and right-handed 
fermions, respectively and ${\rm diag}(m_i,m_j,m_k)$ are mass 
eigenvalues of relevant fermions. 
We assume that the  neutrino masses  
are obtained from the so-called see-saw mechanism 
with huge right-handed Majorana masses ($M_R$) and 
with the Dirac neutrino masses ($M_{\nu_D}$) 
\begin{equation}
M_{\nu}=M_{\nu_D}^T M_R^{-1} M_{\nu_D} \ .
\label{seesaw}
\end{equation}
Generally large neutrino mixing angles may be derivable 
even in the case when the Dirac neutrino mass matrix 
shows strong hierarchical with very small mixing angles 
if $M_R$ is tuned very properly
\footnote{We call such cases  "see-saw enhancement" \cite{enhancement}.}. 
However here we try to find the conditions for reproducing  
the experiments  without 
fine tuning. 

Let us see how the parameters appearing in Eq.~(\ref{generalform}) 
at GUT scale are generally constrained from the present experimental 
neutrino data.
For a moment forget about how to derive the 
parameters of $M_{\nu}$ and just see how the parameter 
regions of $h$ and $\phi$ are 
constrained from the experimental data of 
$\sin^2 2\theta_{\rm atm}, \tan^2 \theta_{\rm sun}$ and the ratio 
of $\Delta m^2_{\rm sun}$ to $\Delta m^2_{\rm atm}$ in terms of 
four parameters $\alpha,\beta,h$ and $\phi$. 
To make numerical calculation more strictly, we must 
take account of the 
contributions from the charged lepton side, 
$U_l$ in Eq.~(\ref{CKMNS}). 
The symmetric charged lepton  mass matrix is written 
in terms of the real matrix $ (\overline{M}_l)_{RL}$ 
and further diagonalized to  $M_l^{{\rm diag.}}$ 
by $ O_l$ \cite{OL}; 
\begin{eqnarray}
 (M_l)_{RL}= P_l^T (\overline{M}_l)_{RL} P_l,  \quad        
 &&  O_l^T\overline{M}_l O_l = M_l^{{\rm diag.}}, \nonumber \\ 
        \rightarrow &&
O_l^T (P_l^T)^{-1} M_l P_l^{-1} O_l \equiv M_l^{{\rm diag.}}.
\end{eqnarray}
We use the following symmetric 
matrix having 2-zeros for $\overline M_l$ , 
\bea
 (\overline{M}_l)_{RL}
\simeq   
\pmatrix{
0                       & \sqrt{m_e m_{\mu}}        & 0 \cr 
\sqrt{{m_e}{m_\mu}}      & m_{\mu}                  & \sqrt{m_e m_{\tau}} \cr 
0 & \sqrt{m_e m_{\tau}} & m_{\tau} \cr
} \ ,
\label{Ml}
\eea
where  $m_e, m_{\mu}, m_{\tau}$ are charged lepton masses at 
$M_{\rm GUT}$ scale. 
Here, we ignore the RGE effect from $M_{\rm GUT}$ to $M_R$ scale considering that 
it almost does not change the values of masses for quarks and leptons. 
On the basis where the 
 charged lepton mass  matrix is diagonalized, 
the neutrino mass matrix at $M_R$ scale is 
obtained from Eq.~(\ref{generalform})
\bea
\tilde M_{\nu}(M_R) = O_l^T (P_l^{-1})^T 
 P_{\nu}^{T}\overline M_{\nu}(M_R)P_{\nu} P_l^{-1} O_l \ ,
\label{netilde}
\eea
\noindent where
\bea
\overline M_{\nu}(M_R)= \left(
\begin{array}{@{\,}ccc@{\,}}
 0               & \beta  & 0   \\
\beta  & e^{i\phi}\alpha   & h \\
 0  & h & 1
\end{array}
\right)m_\nu \ , \quad
 Q \equiv P_{\nu} P_l^{-1}=\left(
\begin{array}{@{\,}ccc@{\,}}
 1              & 0 & 0   \\
 0  & e^{-i\rho}  & 0 \\
 0  & 0 & e^{-i\sigma}
\end{array}
\right) \ .
\eea
In order to compare our calculations with experimental results, 
we need the neutrino mass matrix at $M_Z$ scale, 
which is obtained from the following one-loop RGE's
relation between the neutrino mass matrices at 
$m_Z$ and $M_R$ \cite{Haba}; 
\bea
\tilde M_{\nu}(M_Z) = \left(
\begin{array}{@{\,}ccc@{\,}}
\frac{1}{1-\epsilon_e} & 0 & 0 \\
0 & \frac{1}{1-\epsilon_{\mu}} & 0 \\
0 & 0 & 1 
\end{array}
\right) \tilde M_{\nu}(M_R) 
\left(
\begin{array}{@{\,}ccc@{\,}}
\frac{1}{1-\epsilon_e} & 0 & 0 \\
0 & \frac{1}{1-\epsilon_{\mu}} & 0 \\
0 & 0 & 1 
\end{array}
\right) \ ,
\label{mnuatmz}
\eea
where $\tilde  M_{\nu}$ is the neutrino mass matrix 
on the basis 
where charged lepton matrix is diagonalized 
(see Eq.~(\ref{netilde})).  
The renormalization factors $\epsilon_e$ and $\epsilon_\mu$ 
depend on the ratio of VEV's, $\tan\beta_v$. 
 By using the form of 
Eq.~(\ref{mnuatmz}) we search the region of the 
parameter set $(\alpha, \beta, h, \phi, \sigma, \rho)$ which are allowed 
by experimental data within $3\sigma$:
\begin{eqnarray}
   && 0.82  \le   \sin^2 2\theta_{\rm atm} \ ,\nonumber\\
   &&   0.28 \le   \tan^2 \theta_{\rm sun}   \le   0.64 \ , \nonumber \\
&& 0.73\times 10^{-3}\le \Delta m_{\rm atm}^2 \le 3.8\times 10^{-3} {\rm eV^2}
  \ , \nonumber \\ 
   &&5.4\times 10^{-5}\le \Delta m_{\rm sun}^2 \le   9.5\times 10^{-5}
 {\rm eV^2} \ ,
\label{expbound}
\end{eqnarray}
\noindent which are derived from Eq. (\ref{expangle}).

Figure 1 shows scatter plots of the allowed region of $h, \phi$,  
in which the neutrino experimental results of Eq.~(\ref{expbound}) are
reproduced  by choosing 
the value $\alpha,  \beta, \rho, \sigma$. This shows clearly that 
$h$  cannot be taken too large or too small; $ 0.4\le h \le 3.0 $. 

Also it is interesting that the phase factor $\phi$ 
should not become large, ($ |\phi| \le 70^\circ$).
This may be important since we have never had the 
information of the phases appearing in $M_{\nu}$, 
which is connected to the leptogenesis. 
Let us explore an 
example of the allowed region of the parameters in $(\alpha, \beta)$ 
 plane for the  typical value $h=1.3$.
The allowed region which is consistent with the 
experimental data Eq.~(\ref{expbound}) is shown in Fig.2,
where $\beta$ is allowed to be in both negative and positive.


So far we  have investigated the region of the parameters appearing in the 
neutrino mass matrix of Eq.~(\ref{generalform}) and shown that the parameter region is restricted 
within narrow range by the present experimental data.  
Here we make a comment whether or not a certain GUT model is consistent with 
the bi-large mixing with present neutrino mass differences. 

As an example, let us take a concrete model \cite{BaOba} with the  simplest 
form of right-handed neutrino mass matrix 
with the phase-factored out diagonal matrix, $P_R$, 
\bea
M_R = 
P_R^{T}\left(
\begin{array}{@{\,}ccc@{\,}}
0 & M_1 & 0 \\ 
M_1 & 0 & 0 \\ 
0 & 0 & M_2
\end{array}
\right)
P_R
\equiv m_R 
P_R^{T}\left(
\begin{array}{@{\,}ccc@{\,}}
0 & r & 0 \\ 
r & 0 & 0 \\ 
0 & 0 & 1
\end{array}
\right)P_R\ .
\label{MR}
\eea
This, with the form of 4-zero texture form of $M_{\nu_D}$, 
yields also texture-zero form Eq.~(\ref{generalform}) 
with the phase factored out by 
$(M_{\nu_{D}})_{RL} = P_{\nu_D}^T (\overline{M}_{\nu_D})_{RL} P_{\nu_D}$, 
\begin{eqnarray}
\overline M_{\nu_D}=
\left(
\begin{array}{@{\,}ccc@{\,}}
 0               &a  & 0   \\
a        & b   & c \\
 0  & c & 1
\end{array}
\right) m_{\nu_D}
\rightarrow 
M_{\nu}=
\left(
\begin{array}{@{\,}ccc@{\,}}
 0                 &\frac{ a^2}{r}           & 0   \\
\frac{ a^2}{r}   &2\frac{ ab}{r}+ c^2 & c (\frac{a}{r}+1) \\
 0  & c (\frac{a}{r}+1) & 1
\end{array}
\right) \frac{m_{\nu_D}^2}{m_R},
\label{numass}
\end{eqnarray}
where $a$ and $c$ are real numbers and $b$ is complex one.
We recognize that, 
in order to get  large mixing angle $\theta_{23}$,
the 23 element must be of the same order as the 33 element, 
namely $c (\frac{a}{r}+1)\sim 1$. Since $c\ll 1$, 
$ca/r$ must be of order 1. 
Thus approximate form of $M_{\nu}$ is
\begin{eqnarray}
M_{\nu} \sim 
\left(
\begin{array}{@{\,}ccc@{\,}}
 0               & \beta  & 0   \\
 \beta   &  e^{i\phi}\alpha   &  h \\
 0  &  h & 1
\end{array}
\right) \frac{m_{\nu_D}^2}{m_R}, 
\qquad
\begin{array}{@{\,}c@{\,}}
 \beta \sim  \frac{a^2}{r},    \\
\alpha \sim  \frac{2ab}{r},  \\
 h \sim  \frac{ca}{r}, 
\end{array}
\label{apmnu}
\end{eqnarray}
which clearly shows that none of $a,b,c$ is zero, namely 6-zero texture are 
already excluded by the experimental neutrino data \footnote{
Here, we note that the 6-zero textures for the quark sector have been already 
ruled out by Ramond, Roberts and Ross \cite{RRR}.}.
Now, one example of the symmetric 4-zero texture with  the 
Pati-Salam symmetry \cite{BaOba} provides us with the Dirac neutrino mass 
matrix at the $M_{\rm GUT}$ scale under a simple assumption of the following 
Higgs  configurations:
\bea
M_{U} =
\left(
\begin{array}{@{\,}ccc@{\,}}
 0                 &{\bf 126}           & 0   \\
{\bf 126}           &{\bf 10}            &{\bf 10} \\
 0                 &{\bf 10}            & {\bf 126}
\end{array}\right)
\rightarrow 
\overline M_{\nu_D} \simeq  \left(
\begin{array}{@{\,}ccc@{\,}}
0 & -3\frac{\sqrt{m_u m_c}}{m_t} & 0 \\ 
-3 \frac{\sqrt{m_u m_c}}{m_t} &  e^{i\phi} \frac{m_c}{m_t} &  
\sqrt{\frac{m_u}{ m_t}} \\  0 & \sqrt{\frac{m_u}{ m_t}} & -3 
\end{array}
\right) m_t \ , 
\label{Mnudirac}
\eea
accompanying the phase factor $P_D$ in a same way as Eq.~(\ref{apmnu}).
By comparing Eq.~(\ref{apmnu}) and Eq.~(\ref{Mnudirac})
the parameters $\alpha, \beta$ are expressed in terms of up-quark masses 
at the GUT scale. Thus, we can predict $\alpha, \beta$ from the up-quark 
masses at the GUT scale, $m_u=0.36 \sim 1.28 \mev$, $m_c=209 \sim 300 \mev$, 
$m_t=88 \sim 118 \gev$, which are obtained taking account of RGE's effect to 
the quark masses at the EW scale  \cite{FX}.

 We show the region of $\alpha, \beta$ predicted from 
the  model of Eq. (\ref{Mnudirac}) in figure 3,
where $h=1.3$ and $m_u = 0.36\sim 1.28 ~\mev$ are  taken.
The allowed region predicted from a neutrino mass matrix with 
two zeros of Eq.~(\ref{generalform}) in figure 2 and  the region given by the up-quark masses are separated slighly as seen in figure 3 
if we take the up quark mass at the GUT scale, $m_u=0.36 \sim 1.28 \mev$, 
seriously.

However the light quark masses are ambiguous because of the non-perturbative
QCD effect. Therefore the allowed mass region of
$m_u$ may be enlarged. In the case of  $m_u = 0.36\sim 2.56 ~\mev$,
we obtain the overlapped region around $\alpha\simeq 1.24$ and
$\beta\simeq -0.2$ with $h=1.3$ as seen figure 4. The allowed region on the $\alpha-\beta$ plane in the case of $h=1.3$, 
which is predicted from a  neutrino mass matrix with two zeros 
of  Eq.~(\ref{generalform}).
The allowed region of the  parameters are very narrow as follows:
\bea
 \alpha=1.23\sim 1.24, \ \ \beta= -0.199\sim -0.197, \ \ 
 \phi=  -\frac{\pi}{18}\sim \frac{\pi}{18}, \ \ \rho=\frac{7}{9}\pi \sim  
\frac{11}{9}\pi,
\eea
where $h=1.3$ is taken. 
On the other hand, our results are almost independent of  the phase parameter 
$\sigma$. Hereafter we take $\sigma=0$ in our calculations.
In these parameters, we can predict $U_{e3}$ 
by including the contribution of the charged lepton sector.  
Here we stress that   $U_{e3}$ is  crucial to 
discriminate various models, therefore, we must be careful to 
estimate it by taking account of  the effect of charged lepton mixings 
as well as CP violating phases. Our formula has already included these
contributions. By taking the overlapped region of $\alpha$ and $\beta$
in figure 4, we present the prediction of  
 $|U_{e3}|$, $J_{CP}$ and $<m_{ee}>$ as follows:

\bea
 |U_{e3}| =0.010 -0.048\ , \quad |J_{CP}| \leq 9.6\times 10^{-3}\ ,\quad
  |<m_{ee}>|\simeq 0.0027 \rm eV \ ,
\eea
where  $<m_{ee}>$ is the effective neutrino mass in the neutrinoless
double beta decay.
We hope $ |U_{e3}|$ can be checked by the neutrino experiments in near future.
 Since  the overlapped region of $\alpha$ and $\beta$ is restricted
in the narrow region, we can predict
 a set of typical values of neutrino masses and mixings at $h=1.3$ 
as follows;
\bea
&&\sin^2 2\theta_{\mu \tau} \sim 0.98, \qquad 
\tan^2 \theta_{\mu e} \sim 0.28, \nonumber \\
&& m_{\nu_{3}} \sim 0.062~\rm eV, \qquad
m_{\nu_{2}} \sim 0.0075~\rm eV, \qquad
m_{\nu_1} \sim 0.0014~\rm eV,
\eea
with $m_R= 3.0\times 10^{15}~\gev$ and $r\times m_R=1.0\times  10^{9}~\gev$, 
which correspond to the Majorana mass for the third generation and 
those of the second and first generations, respectively.
On the other hand, $m_u \simeq 2.56 ~\mev$ should be allowed at the GUT scale.
Now that our neutrino mass matrix is determined almost uniquely 
from the up-quark masses at GUT scale, we can make the prediction of 
leptogenesis once we fix the CP violating phases. 
Interesting enough is that our form of $M_R$ of Eq.~(\ref{MR}) 
yields naturally two degenerate Majorana masses with mass 
$r\times m_R\sim 10^{9}$ GeV. 
In such case the leptogenesis is enhanced by the so-called "crossing effect" 
\cite{smirnovcrossing}, 
which are now under calculation by 
Bando, Kaneko, Obara and Tanimoto\cite{leptogenesis}. 

In conclusion we have shown that, in order to be compatible with the 
present neutrino experiments, the parameters of a neutrino mass 
matrix with two zeros in Eq.~(\ref{generalform}) are constrained to 
a small region.  
Also, we have seen that the 4-zero texture with Pati-Salam symmetry 
restricts the above prameter region to a very narrow region indicated 
in Figure 4, enlarging the values of up quark mass at the GUT scale.  
Both parameter regions should be compared in detail, which will be
published elsewhere in the near future.
The precision mesurements, especially, for the solar neutrino 
mixing angle and the mass squared differences will check if such 
a texture of geometric form with Pati-Salam symmetry 
is realized in Nature in the near future.




\section*{Acknowledgements}
This collaboration has been encouraged by the stimulating discussion in the 
Summer Institutes 2003. 
We would like to thank to  A. Sugamoto, T. Kugo and N. Okamura and 
T.  Kurimoto
who encouraged  us very much. 
 We are also indebted to SI2003 by many
helpful discussions. 
M. Bando  and M. Tanimoto  are  supported in part by
the Grant-in Aid for Scientific Research No.12047225 and 12047220. 
%
%
%
%
%
%
\newpage

\newpage
\begin{figure}
\epsfxsize=17.0 cm
\centerline{\epsfbox{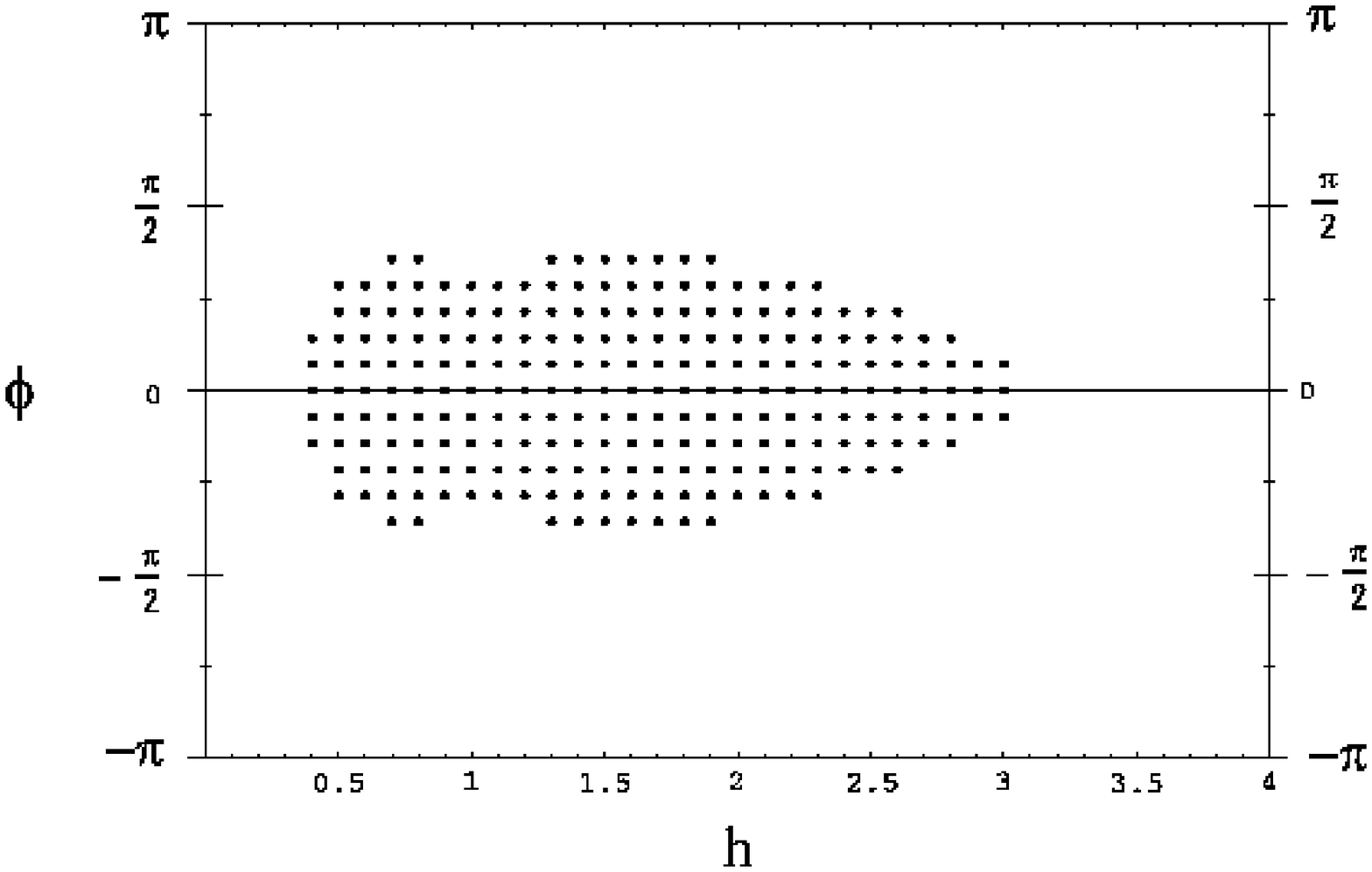}}
\caption{The scatter plots of the allowed region on the $h-\phi$ plane. }
\end{figure}
\begin{figure}
\epsfxsize=14.0 cm
\centerline{\epsfbox{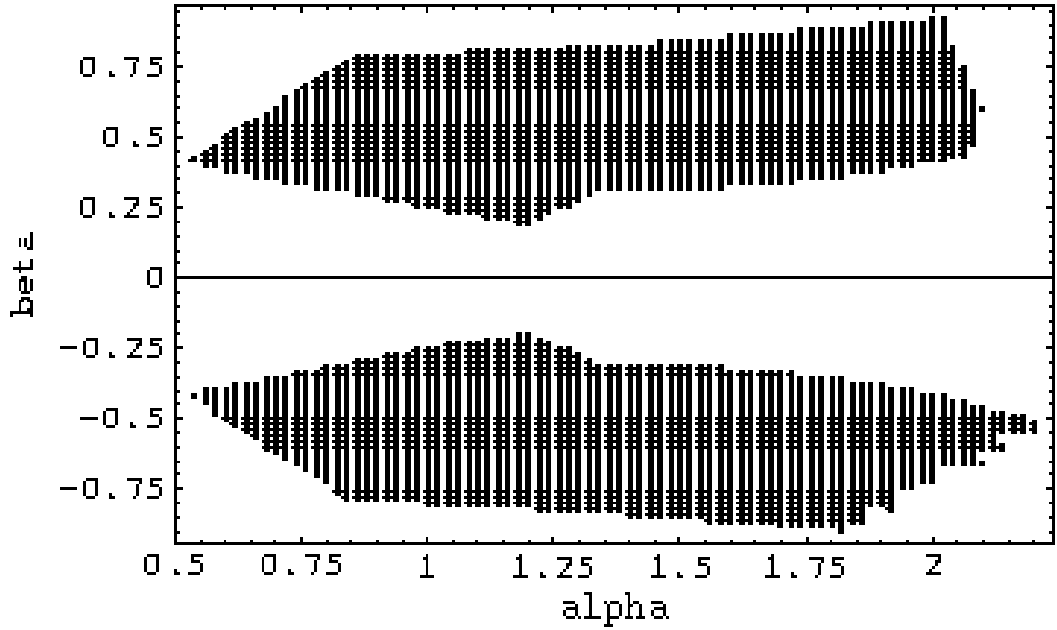}}
\caption{The allowed region on the $\alpha-\beta$ plane in the case of 
$h=1.3$,  which is predicted from a  neutrino mass matrix with two zeros of 
 Eq.~(\ref{generalform}).}
\end{figure}
\begin{figure}
\epsfxsize=12.0 cm
\centerline{\epsfbox{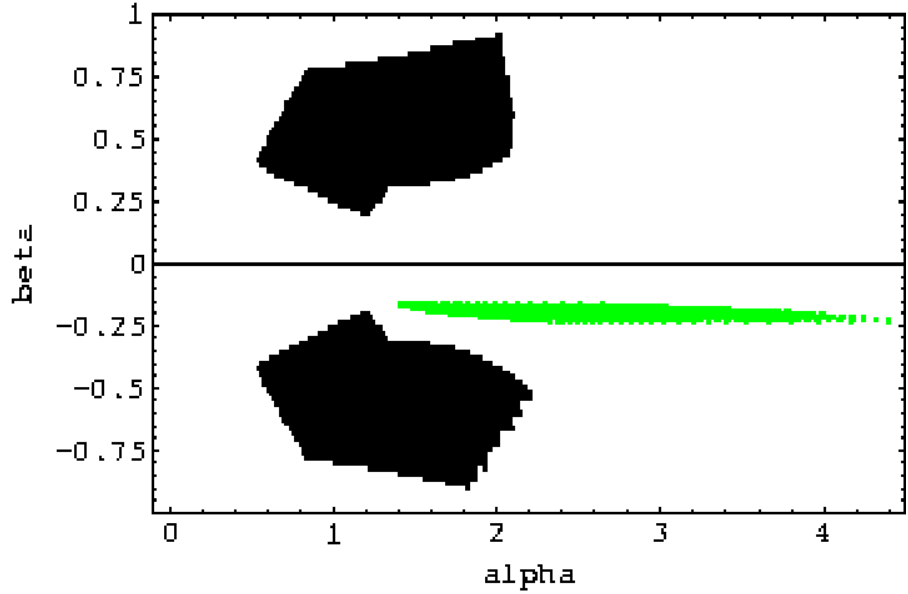}}
\caption{The predicted region (gray region) of the $\alpha-\beta$ plane in the GUT model, where $h=1.3$ and $m_u = 0.36\sim 1.28 ~\mev$ are taken.
The black region is the experimentally allowed region 
 predicted from a neutrino mass matrix with two zeros of 
Eq.~(\ref{generalform}).}
\end{figure}
\begin{figure}
\epsfxsize=12.0 cm
\centerline{\epsfbox{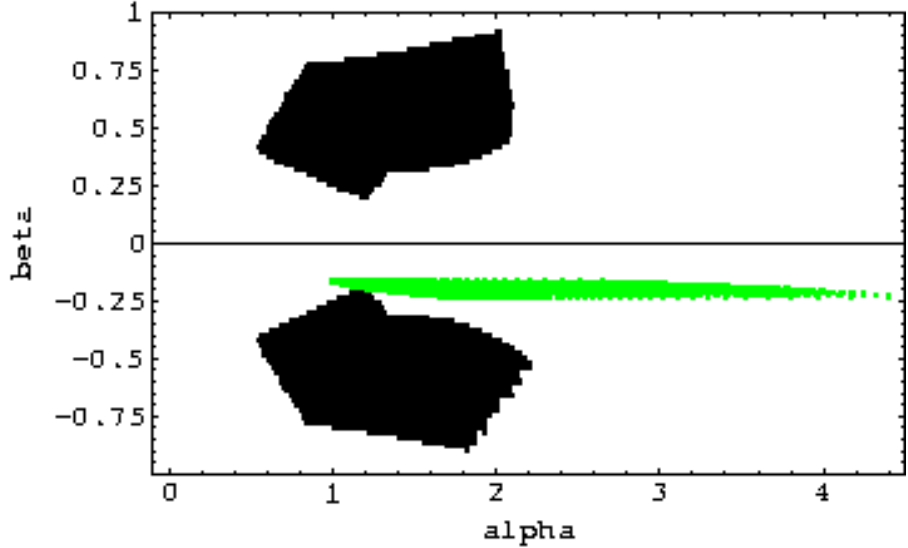}}
\caption{The predicted region (gray region) of the $\alpha-\beta$ plane, 
in which  $h=1.3$ and $m_u = 0.36\sim 2.56 ~\mev$ are taken.  
The black region is the experimentally allowed region  predicted from a neutrino mass matrix with two zeros of 
Eq.~(\ref{generalform}). There is the overlapped region 
around $\alpha\simeq1.24$ and $\beta\simeq-0.2$. }
\end{figure}

\end{document}